# On some properties of medians, percentiles, baselines, and thresholds in empirical bibliometric analysis*

Vladimir Pislyakov

Library, HSE University, 11 Pokrovsky blvd. Moscow, 109028, Russia

*E-mail address:* pislyakov@hse.ru

**Highlights**

- We test common and "self-explaining" ranking terminology in bibliometrics
- Due to different reasons, "medians" are often not halves, "quartiles" are not quarters, percentiles are not hundredths
- Almost half of all world papers are published in Q1 journals
- The effect of whole/fractional counting on normalized indicators is studied
- We propose more complicated but more well-founded methods for addressing the "borderline" highly cited papers

**Abstract**

One of the most useful and correct methodological approaches in bibliometrics is ranking. In the context of highly skewed bibliometric distributions and severe distortions caused by outliers, it is often the preferable way of analysis. Ranking methodology strictly implies that "oranges should be compared with oranges, apples with apples". We should make a "like with like" comparison. Ranks in different fields show how a unit under study is compared to others in its field. But do we always apply an "apples approach" appropriately? Is median really a 50%, quartile a 25%, 10th percentile a 10%? The paper considers theoretical definitions of such terms compared to their real sense in the course of bibliometric research. It is found that in many empirical cases quartiles are not quarters, medians are not halves, world baselines are not unity, and integer thresholds lead to inequality of performance evaluation in different science fields.

## 1. Introduction

In bibliometrics, it is often more convenient and robust to use ordinal rather than cardinal methods. Bibliometric distributions are usually non-Gaussian and highly skewed, so direct comparison of values and metrics, as well as their averages, is sometimes misleading. Even the most carefully elaborated and fully normalized cardinal indicators may suffer from outliers. That is why it is often better to use medians instead of means, ranks instead of values.

A ranking approach means that we put aside absolute values of bibliometric indicators and consider only relative positions—which of the entities is higher, and which is lower. Such a series of "tournaments", "paired comparisons" (Kóczy & Strobel, 2010; David, 1963) results in an ordered ranking. This attitude often helps to smooth differences in highly skewed bibliometric distributions. For example, a long-standing champion of impact factor (IF), *CA: A Cancer Journal for Clinicians*, received a 2020 IF of 508.7. Its nearest peer, in the "Oncology" category, *Nature Reviews Clinical Oncology*, has an IF of 66.7. Does it mean that the former journal is more than 7.6 times "better" than the latter? Probably not. If we use a journal ranking approach, it just asserts that they have the first and second positions in the "Oncology" discipline.

With some reservations, ranking methods enable cross-disciplinary comparison if we use not the ranks themselves, but the relative position of the journal in the ranking of its subject category. This exercise may be done with various levels of granularity. The simplest and roughest is the "higher than median" vs. "lower than median" division. Next, quartiles, division of the ranking into four equal intervals, are widely known. Sometimes, deciles, i.e. 90th/80th/70th, etc. percentiles are used. The most precise is the usage of the exact percentile which is the relative place of a journal in the ordered ranking of its discipline. It shows what proportion of journals is "weaker" than the publication in focus. For example, *Journal of Informetrics* is ranked 18th by IF in the discipline *Information Science & Library Science*, which contains 86 titles. It has $(86-(18-0.5))/86 = 79.7\%$ percentile (the formula includes "half of a journal itself"). This means that *JoI* has an IF higher than about 80% of journals of the same science field. Now this indicator is published in each year's JCR edition for all journals, it is called "Average JIF Percentile", where "average" means averaging of the percentile if a journal is assigned to several subject categories. However, this principle was not newly introduced to bibliometrics by Clarivate databases, see for example, conceptually close "normalized journal position", defined by Bordons and Barrigón (1992).

In fact, the choice of granularity level in ranking is quite an ambivalent task. On the one hand, the higher the resolution of our method, the more journals we may categorize and determine which of them has better metrics. On the other hand, a rougher division of the journal sets complies with our intuitive idea that journals of "almost equal prominence" exist (cf. Subochev, Aleskerov, & Pislyakov, 2018, p. 425).

For the purposes of benchmarking, cross-disciplinary comparison, or implementation of ranking indicators, such concepts as "thresholds", "percentiles", "baselines" are often used (e.g. Shaw, 1985; Schreiber, 2013; Colliander & Ahlgren, 2011). Generally, they have names which speak for themselves, such as "median", "quartile", "10th percentile", "world baseline", etc. However, sometimes the labels are deceptive, and it is important always to keep in mind the empirical sense of these variables when one implements them for real data and databases.

This paper explores some important peculiarities and nuances of these concepts and demonstrates their properties on a theoretical as well as empirical basis. The purpose of the present work is to reinforce the correct framework for the usage of the above-mentioned terms and draw additional attention to the implementation of this conceptual cloud in a proper way, especially for bibliometricians just beginning their sojourn. This means, we aim to (a) demonstrate methodological and terminological traps in basic bibliometrics; (b) indicate ways to amend the approaches when it is possible for a researcher; (c) call on the assistance of major database providers when it is not.

We will use a specific package of bibliometric instruments for demonstration purposes. It is one of the most well-developed analytical databases, InCites from Clarivate company, including Journal Citation Reports (JCR) and Essential Science Indicators (ESI) sub-databases.[1] They contain all the units necessary for this study: journal impact factors, normalized indicators, and highly cited papers. However, all concerns are in no way exclusive for these instruments and are appropriate for many bibliometric contexts when other data sources are used.

The structure of the paper is as follows. Sections 2–4 consider different cases when percentile or world baseline approaches contains hidden traps. In section 2, the evenness of journals' (2.1) and papers' (2.2) distribution across journal quartiles is tested. Section 3 explores concepts of the "world baseline" (3.1) and the "reference group baseline" (3.2) in subject normalization methodologies, when these supposed to be baselines may appear misleading in the process of benchmarking. Highly cited papers are explored in section 4, with their problems of multi-authorship/fractional counting (4.1) and integer thresholds/quantization (4.2). Sub-section 4.3 contains practical suggestions on determining highly cited papers in borderline cases, while empirical data analysis from ESI highly cited papers database is presented in sub-chapter 4.4. Finally, section 5 sums up the possible implications of this study, and section 6 concludes the paper.

---

[1] Vide: https://clarivate.com/webofsciencegroup/solutions/incites/
https://clarivate.com/webofsciencegroup/solutions/journal-citation-reports/
https://clarivate.com/webofsciencegroup/solutions/essential-science-indicators/

## 2. Quartiles

Journal ranking by impact factor quartiles is a widespread practice in bibliometrics as well as in science administration (e.g. Lisitskaya et al., 2018; Miranda & Garcia-Carpintero, 2019; Orbay, Miranda, & Orbay, 2020). The most prominent advantage of rank indicators is their capacity to resolve cross-disciplinary collisions in a quite satisfactory way.[2] Quartiles are defined within each journal's scientific discipline, allowing bibliometricians to compare quartile numbers across them.

It seems that the term "quartile" speaks for itself. It implies 25%-25%-25%-25% division of the journal pool. However, when we talk about journal quartiles, we encounter two kinds of traps. Those associated with the number of journals and those related to the number of articles/papers.

### 2.1 Number of journals

It is obvious that, within each specific discipline, the Q1-Q4 proportions of journals are exactly 25% only if their total number N is divisible by 4. What happens if it is not? Then (according to JCR internal algorithms), in a journal list sorted by impact factor, Q1 journals are those from #1 to [N/4], Q2 from [N/4]+1 to [N/2], Q3 from [N/2]+1 to [3N/4] and Q4 from [3N/4]+1 to N. (Brackets here denote rounding down to the nearest integer.) Fig. 1 shows four cases for disciplines with 16–19 journals in each, where the numbers of journals have different remainders modulo 4.

What does it mean? It means that, if we assume that the number of journals in a discipline is uniformly (i.e. randomly) distributed across remainders modulo 4, we have a relative surplus of Q2 and, especially, Q4 journals across the whole database and a relative lack of Q3 and, especially, Q1 journals. We may obtain some probabilistic estimation. There are 236 subject categories in JCR. In 3/4 of them there will be a surplus of one Q4 journal compared to Q1, in 1/2 a surplus of one Q2 journal, and in 1/4 of cases there will be one more Q3 journal than the number of Q1 titles. This means, approximately, +118/+59/+177 journals of 2/3/4 quartiles compared to Q1, respectively (cf. Fig. 1). Is it a lot or a little? We have a base of ~12,100 journals in JCR. There are 118 more Q2 journals than Q1, Q2=Q1+118. Similarly, Q3=Q1+59, Q4=Q1+177. Adding up, we get approximations Q1=2937, Q2=3054, Q3=2996 and Q4=3113. This is a deviation from the average (3025) within about 2.9%.[3]

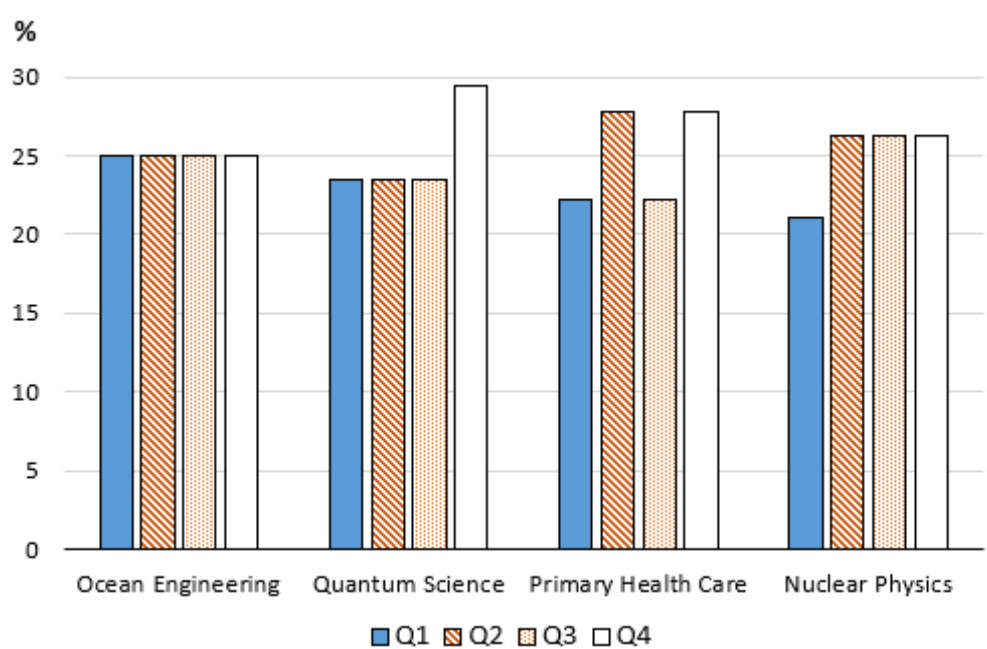

**Fig. 1.** Journal distribution by quartiles, four disciplines (JCR-2020).

Thus, if we gather Q1-Q4 journals across the whole database following a discipline-by-discipline method, we get less than 25% of Q1 and Q3 journals and more than 25% of Q2 and Q4 titles (Fig. 1). Surprisingly, this is not the case if we consider InCites as a whole (Fig. 2).

There are 3284 Q1 journals out of 12113 titles with an impact factor in InCites-2020. It is 27.1%. In Q1+Q2 (which is the "median"), there are 52.6% of journals. What is the explanation? The main factor is that many JCR journals are attributed to several (mostly two or three) subject categories. When the end-to-end search is made in InCites, quartiles are attributed according to "the highest one" principle. As the InCites Handbook explains, "InCites uses the best quartile for journals that appear in multiple Web of Science Research Areas" (Clarivate, 2018, p. 9).[4]

---

[3] This is an approximate mathematical estimation. We cannot coherently compare it with empirical JCR data because of deviations from the uniformity of journal distributions across subject categories and cases of several subject attributions to one journal (see further).

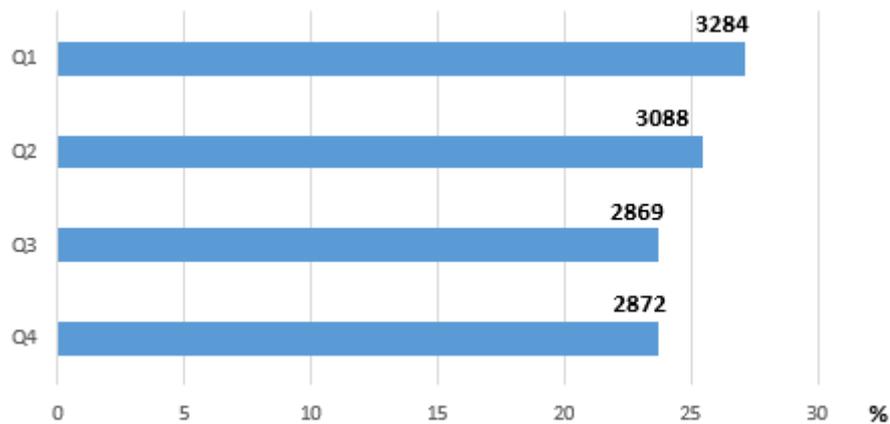

**Fig. 2.** Journal distribution by quartiles, the whole InCites-2020
(only journals with impact factor/quartile assigned are considered).

Both effects should be kept in mind when we talk about "journals of the first quartile". Their number is usually *less* than one in four if we speak of a particular science category. It is *more* than one in four if we speak of the database as a whole. In the latter case it means, in absolute numbers, that now there are 412 more Q1 than Q4 journals.

*2.2 Number of papers*

If journal distribution is not uniform across quartiles, the papers/articles are even further from that. Whether it is by accident or not, on average, higher quartiles are occupied by more voluminous journals (in terms of the annual number of papers). This leads to an extraordinary advantage in the number of papers in Q1 journals (Fig. 3). This, in a sense, contradicts with Antonoyiannakis (2018), who considers all journals without discipline division and comes to the conclusion that bigger journals are penalized in their impact factor.

It is an astonishing fact that, as Fig. 3 shows, nearly *every 2nd paper* is published in the best quartile journal (in one of the disciplines). Consequently, this mark of excellence should be taken *cum grano salis*. Inside Q1+Q2 (supposedly "median"), there are 73.3% of papers.

---

[4] There is another more technical factor distorting the picture in InCites. Some titles which are periodicals in JCR and obtain their IF there, are considered as volumes of book series in InCites. Consequently, they do not have neither IF nor quartile in this database (examples: *Annual Review of Virology*, *Advances in Virus Research*). They still "occupy" the place in JCR ranking resulting in decrease of the number of titles of their quartile in InCites.

Meanwhile, only one in nine papers appears in the Q4 journal. In other words, eight out of nine random papers are published in a journal which belongs to Q1, Q2 or Q3 in one of the subject categories in InCites. It may be a Q4 journal at the same time in some other category, but that is another question.

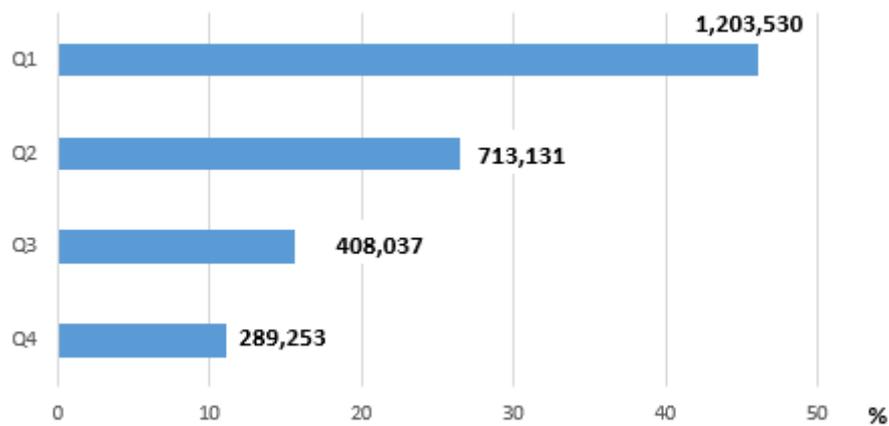

**Fig. 3.** Share of papers published in journals of different quartiles,
the whole InCites-2020 (only journals with impact factor/quartile assigned are considered).

Moreover, this unevenness of papers' distribution across quartiles may be different in different research areas. This leads to inequality of chances to publish a high quartile paper depending on the field of science where a researcher or a team works.

## 3. Baselines in subject normalization

A comparison of scientific entities (authors, laboratories, organizations, countries, journals) operating in different science fields is a well-known problem in bibliometrics. Average levels of citations in various scientific disciplines differ drastically. For example, the median journal impact factor in 2020 for cell biology is 4.51, for oceanography 2.33, and for mathematics 0.96. This inequality in average citation indicators has nothing to do with the scientific level and excellence of organizations, authors, etc. working in these fields. It is related to the general "citation practices" in different disciplines, first of all, the average length of the references lists, chronological distribution of the bibliography sources, proportion of review articles and weight of the journal literature in the field.

### 3.1. World baselines

As was mentioned, general discipline factors do not represent the "scientific quality" of the entity under examination. One should try to use some normalization to overcome them. For this purpose, one of the most complex and smart indicators is introduced in InCites. It is the Category Normalized Citation Impact, CNCI. "Category" here means subject category, science field—according to one of the 20+ global and regional research area schemas as an available option (for example, WoS categories, OECD, UK REF). This metric allows comparison between papers of different age, science field, and document type (Article, Review, etc.). In brief, it is calculated as a ratio of actual citations received by a paper and its "expected" value, that is an average citedness of a paper of the same year of publication, scientific discipline, and document type.

What is essential here, is the existence of a *world baseline*, when we may say whether CNCI of some unit is "better" or "weaker" than one should expect, taking into account the whole context. Usually, it is believed that if the CNCI > 1 for a single paper, this means that it is cited above average, when calculated for "the same" world papers. If the CNCI < 1, this paper is "weaker" than its counterparts. Moreover, more generalized statements, such as "a CNCI value of two is considered twice world average" (Clarivate, 2018, p. 14), are often declared. Clarivate is in no way the first who introduced subject normalization, CNCI has its predecessors: $R_w$ by Vinkler (1986),[5] NMCR by Braun and Glänzel (1990), CPP/FCSm by de Bruin et al. (1993) and Moed, de Bruin, & Van Leeuwen (1995).

Let us consider how CNCI is calculated for a set of papers coming from different disciplines. But let's start with the logical assumption that CNCI *for the whole world* (all papers in the world indexed in InCites) should be equal to 1. The world as a whole is, by definition, *a world average*. Its relative citation indicator should equal unity, simply because the reference group coincides with the evaluated group.

However, there are some peculiarities in CNCI methodology which lead to (a slight) violation of this basic principle. The problem is, again, with the papers attributed to several disciplines at once (as soon as the journals, where they are published, are listed under more than one subject category). For such papers, their CNCI is calculated as an average of CNCIs within each discipline (Clarivate, 2018, p. 14). Let some paper of document type $t$ be published in a year $y$, in a journal assigned, for example, to three different science fields $f_1$, $f_2$, and $f_3$. Using notations of Clarivate (2018), with small changes, CNCI for this paper is

$$CNCI = \frac{1}{3}\left(\frac{c}{e_{f_1yt}} + \frac{c}{e_{f_2yt}} + \frac{c}{e_{f_3yt}}\right) \qquad (1),$$

where $c$ is an actual number of citations received by the paper, $e_{fyt}$ is its expected (average) value for a science field $f$, year of publication $y$, document type $t$. (NB: in the InCites Handbook there is a misprint in this equation, CNCI is designated as NCI, a vestige from the previous edition.)

---

[5] Relational charts of observed vs. expected mean citation rates by Schubert and Braun (1986), appeared a bit earlier than Vinkler's work, however they link "expected" to mean *journal* value, thus more anticipating another InCites indicator, Journal Normalized Citation Impact (JNCI). At the same time, Vinkler (1986), used a reduction factor of 0.85 (thus trying to allow for self- and "indoor" citations), which did not have any further development in bibliometrics.

At last, the CNCI for a set of papers is calculated as the simple arithmetic mean of CNCIs of all documents in the set (Clarivate, 2018, p. 14).

Now let us consider an extreme imaginary example. Suppose the whole world literature consists of only two papers, P1 and P2. Both are of the same document type and were published in the same year. Paper P1 belongs to only one discipline, A, and received 1 citation. Paper P2 is assigned to two scientific disciplines, A and B, and received 2 citations (Fig. 4).

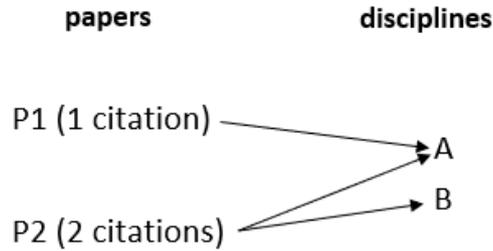

**Fig. 4. Fictional example for calculating world CNCI ("global baseline").**

In this case, the expected number of citations (baseline) for the discipline A would be

$$Baseline_A = (c_{P1} + c_{P2})/2 = (1 + 2)/2 = 1.5 \qquad (2)$$

($c_p$ here is the number of citations received by paper P). For discipline B:

$$Baseline_B = c_{P2}/1 = 2/1 = 2 \qquad (3).$$

Then CNCI values for each paper are

$$CNCI_{P1} = c_{P1}/Baseline_A = 1/1.5 = 2/3,$$

$$CNCI_{P2} = (c_{P2}/Baseline_A + c_{P2}/Baseline_B)/2 = (2/1.5 + 2/2)/2 = 7/6.$$

For "the whole world", that is a set of both papers P1 and P2, we get CNCI:

$$CNCI_{\text{Global}} = (CNCI_{P1} + CNCI_{P2})/2 = (2/3 + 7/6)/2 = 11/12 < 1 \qquad (4).$$

Simple repetition of these calculations for the case when P1 receives two citations and P2 only one would lead us to CNCI of the world 13/12, that is > 1.

This strange effect, that the normalized baseline of the whole world may be greater or lesser than unity, should be kept in mind when we evaluate scholars or assess organizations. In fact, this should be considered a problem for the whole CNCI framework, as soon as the only undoubtable benchmark fails. We should remark that the problem is known for bibliometricians who develop InCites, as they insert a note, however without any demonstration or reflection: "A quirk of the way baselines are calculated [...] and the way CNCI is calculated [...] results in the CNCI of the world not being equal to one exactly" (Clarivate, 2018, p. 14). The same predicament, but for country/organization/author level, is studied in detail by Waltman and van Eck (2015).

The similar unusual result may be observed, not only for the whole world, but for each subset of the documents of the same type published in the same year. Theoretically, every such slice should have a total CNCI equal to 1. However, now, this is not the case in InCites, according to the same logic of equations (2)–(4).

Actually, among 20+ research area schemas of the InCites, there is only one where this problem does not occur. It is the Essential Science Indicators (ESI) science field division. This is the only classification where each journal is attributed strictly to one (broad) discipline. There is no "multiple subject categories" issue there. The empirical data confirm it perfectly. Fig. 5 shows Global CNCIs (baselines) in the case of a standard Web of Science journal categories' classification, which allows multiple attribution, and for ESI unique-only categorization. For the latter, the global indicator equals exactly 1. For the Web of Science schema, global CNCI is 0.98. By varying the time period and document types, we succeeded in getting values from 0.98 to 1.00, while the ESI indicator never changed. Additionally, Table 1 shows the range of Global CNCIs for four global and three regional subject schemas,

for different timespans starting between 2015–2019 and ending in 2019 (2015–2019, 2016–2019, and so on; 2020 is not considered to avoid outliers on a small timespan).

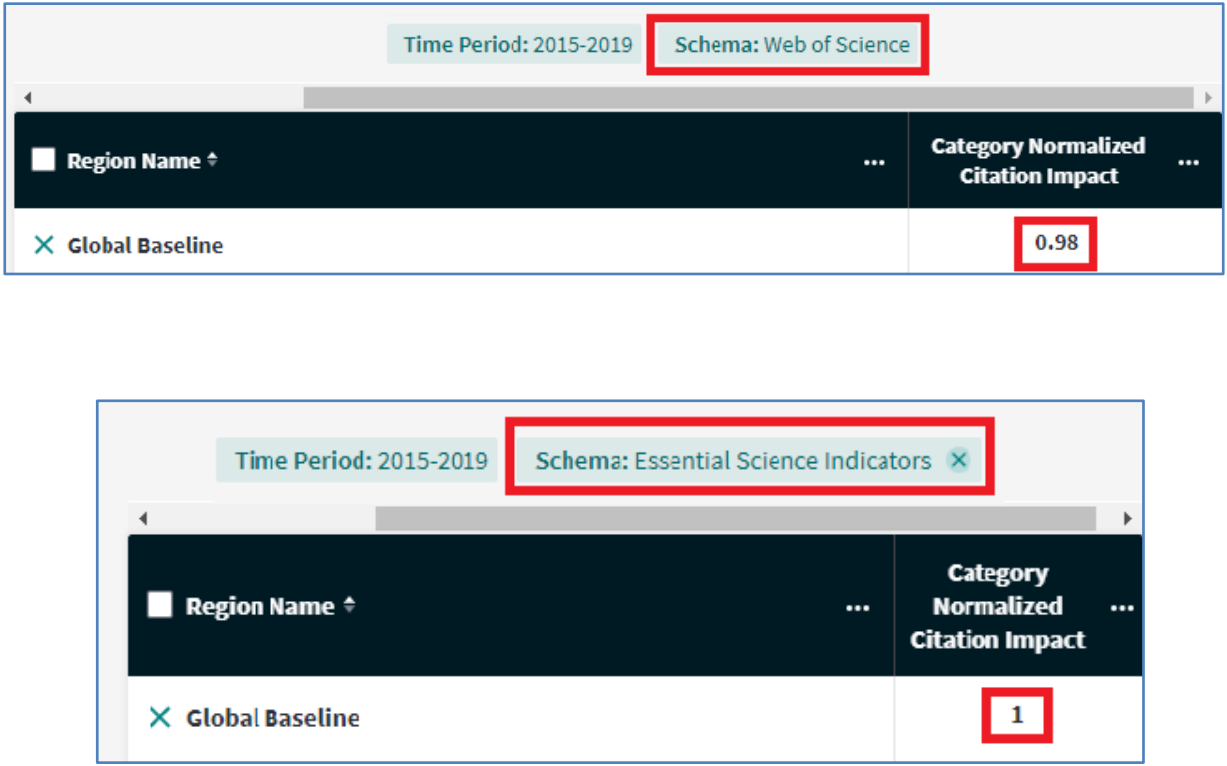

**Fig. 5.** Global CNCI for two different research area schemas: Web of Science categories (top) and Essential Science Indicators fields (bottom). Source: InCites (edited screenshots).

Another observation follows. If we extrapolate the example of Fig. 4 and result of equation (4), we may suppose that if more citations are received by papers attributed to several subjects, then global CNCI, calculated by InCites methodology, is smaller than unity. On the contrary, if single-focused papers are cited more, then global CNCI should be greater than 1. As Table 1 shows, usually, the global InCites baseline is less than 1, no matter which multi-classification schema is used.[6] It is known that, now, many discoveries occur when there is a combination of ideas coming from different science sectors. Such discoveries and new technological ideas should be abundantly cited (e.g. Kwon et al., 2019). Our observation is also in agreement with several recent studies, which argue that interdisciplinary knowledge is better cited than more narrow research (Larivière, Haustein, & Börner, 2015; Petersen et al., 2018; Chen et al., 2021), at least if cross-disciplinarity is not excessive (Yegros-Yegros,

---

[6] We found only one classification with its two versions, which yielded a global baseline greater than 1. It is the State Council Academic Degree Committee (SCADC) of China schema, "SCADC Subject 97 narrow" (1.03–1.05) and "SCADC Subject 13 broad" (1.01–1.02).

Rafols, & D'Este, 2015). This may be the clue to common case of Table 1 when global CNCI turns out to be smaller than unity.

**Table 1**
Global baselines (CNCI) with different timespans and different document types. Whole InCites without ESCI, different research schemas. Publication window, starting from 2015/16/17/18/19, ending in 2019. Document types tested: only Articles; only Reviews. Description of schemas: (Clarivate, 2018).

| Research schema | Range of global CNCI |
|---|---|
| Web of Science | 0.98–1.00 |
| Essential Science Indicators | 1.00 |
| OECD | 0.97–0.99 |
| GIPP | 0.98–0.99 |
| UK REF 2014 | 0.98–0.99 |
| ANVUR | 0.97–0.98 |
| FAPESP | 0.97–0.99 |

Is there a way to coherently resolve this issue, to overcome this weakness in CNCI methodology, and get a world baseline equal to unity? What changes to methodology should be introduced? There are two options. The first keeps in place "whole counting" of publications, while the citations should be equally distributed between all the disciplines a paper belongs to. Moreover, we should return to the previous practice of InCites ("InCites-1") before 2015. Originally CNCI (called, at that time, NCI) was calculated as the ratio of the sum of all received citations by the papers, to the sum of their expected values. This is a difference between the "ratio of the averages" approach compared to the "average of the ratios", which is now used in InCites and, consequently, in Eq. (4).

Let us take discipline A from Fig. 4. It received 1 citation from P1 and half of the citations from P2 (as soon as P2 is assigned to 2 disciplines). The sum is 2 from 2 papers, the expected citedness of papers attributed to A is 1. Discipline B received half of the citations of P2, expected value is 1. The overall expected rate is 1*2 (two papers in A) + 1*1 (one paper in B) = 3 citations. As we see, the whole set of papers received 3 citations and NCI of the system is received / expected = 3 / 3 = 1. This will hold true for any closed universe of publications.

There was a deep and fundamental controversy between the "ratio of the averages" vs. the "average of the ratios" approaches in normalization. The trail of this dispute may be found in the flood of letters to the *Journal of Informetrics* in the early 2010s (Opthof & Leydesdorff, 2010; Van Raan et al., 2010; Bornmann, 2010; Gingras & Larivière, 2011; Bornmann & Mutz, 2011; Leydesdorff & Opthof, 2011). In the end, this led to a change in CWTS methodology and transition from the "ratio of the averages" approach to the "average of the ratios". The logic and theoretical justification for this change is thoroughly explained by Waltman et al. (2011b), with more ambiguous results of empirical tests (Waltman et al., 2011a).

However, there is no need to change the way of averaging if we move to the full-fledged fractional counting method. Then discipline A receives 1+2/2=2 citations from 1+1/2=1.5 papers (P2 as a paper with its citations is distributed across its two disciplines). Therefore, its own expected baseline is 2/1.5 = 4/3. For discipline B, the same calculations result in expected citations (baseline) of 2: one citation from half of paper. Then

$$CNCI_{P1} = c_{P1}/Baseline_A = 1/(4/3) = 3/4,$$

$$CNCI_{P2} = (c_{P2}/Baseline_A + c_{P2}/Baseline_B)/2 = (2/(4/3) + 2/2)/2 = 5/4.$$

We may see that global CNCI in this case equals unity: (3/4+5/4)/2. These calculations use the "average of the ratios" method. An additional advantage of this fractional method is that it makes possible a proper field normalization, which cannot be achieved with full counting (Waltman & van Eck, 2015).

Unfortunately, at present, InCites does not offer fractional counting methodology at any level of analysis. Only first-author and corresponding-author analysis were added in April 2020. Some Clarivate-affiliated researchers privately create their own scripts which calculate fractional indicators based on the WoS API (Kasyanov, 2021). It should also be mentioned that some government initiatives concerning science policy and evaluation of organizations already introduce fractional counting of publications and citations. For example, in Russia, in 2020, a new indicator for reporting to the Ministry of Science and Higher Education, "Complex grade of publication performance", was piloted. An author-level fractional counting was used for its calculation (Ministry of Science and Higher Education of the Russian Federation, 2020). Also, a brand-new program of academic leadership "Priority-2030" for Russian universities, in its call for participation, prescribes the use of fractional attribution of papers to scientists/organizations (Ministry of Science and Higher Education of the Russian Federation, 2021).

*3.2. Relative comparison*

Another subtle issue with CNCI and similar indicators is that their "reference group" is *the whole world*. If we know that the CNCI of some unit (author/organization, for example) is more/less than unity, it means (with reservations of the previous sub-section) that this unit outperforms/underperforms an average world author/organization with the same output structure.

But what if we would like to know how the unit performs within some larger scientific entity, but not the whole world? For example, how a lab looks in comparison to the whole institute or an organization in comparison to its country? Can we use the relation between the CNCI values of these two entities, the unit and its sub-unit, for example, their ratio?

Unfortunately, no. Let us consider the performance of the whole of Russia and one Russian institute, The Boreskov Institute of Catalysis (city of Novosibirsk). The CNCI of all Russian output in 2015–2019 is 0.79, while this institute's CNCI is 0.70. One may suppose that the Institute of Catalysis performs worse in comparison to the background of its country. However, the specialization of this scientific unit is chemistry. And if we look at the normalized indicator of Russian science for chemistry, we find that it equals 0.54. It means that the Boreskov Institute of Catalysis outperforms an average Russian institution of its scientific profile, while a higher CNCI for whole of Russia comes from other fields, in particular, clinical medicine and neuroscience.

This is a simple example of highly specialized organization. In general, to give a correct answer, we should calculate for the sub-unit in question an indicator similar to CNCI, where the reference group will be not the whole world, but the larger unit. We should compute expected values as the average number of citations of papers of the same year, field and type published by a reference entity (larger unit). Next, we take a ratio of number of real ("observed") citations to this expected value for each paper of the sub-unit and then average these ratios over all the sub-unit's publications. Only in such a way we may get a correct relative citation indicator, for the smaller entity against the larger one comprising it. No direct arithmetic manipulations with CNCIs may help. Here CNCI serves as a "false common baseline", which compares units with the whole world but does not permit to compare them with one another.

## 4. Thresholds for highly cited papers

Another bibliometric domain where thresholds are used is research on highly cited papers. If it is organized according to the definition of the InCites/ESI database, the highly cited document is a paper which belongs to the top 1% most cited papers published in the same science discipline and the same year. How should one use this percentile and are there really 1% of the highly cited papers?

### 4.1. Multi-authorships

First of all, the 1% level may be used as a reference point only for the benchmarking of journals. If 1% of the journals' papers become highly cited, according to the definition above, then this is a journal "of average prominence", in line with the whole world literature. If we try to benchmark authors, departments, organizations, or countries, we should use the 1% criterion only for the *fractional counting* method, that is, when a paper is attributed to a scientific entity in a proportion of its share in the list of co-authors. This is especially important for highly cited papers. They are usually written in large co-authorships.

For example, Pislyakov and Shukshina (2014) found that 927 of the papers by Russian scientists, published in 2000–2009, became highly cited. That is about 0.37% of the total Russian output, three times below the average. However, these are results for a whole counting method. In fact, the average share of Russian authors (i.e. the proportion of authors with Russian affiliation) is about 30% in highly cited papers (Pislyakov & Shukshina, 2014). It would be wrong to conclude that only 0.37*0.3 = 0.11% of Russian output becomes highly cited as soon as total number of Russian papers should also be counted fractionally. This is hard to calculate, as there is no proper publicly available bibliometric instrument. However, we may suppose that the average share of Russian authors is usually higher than 30% in the total output. There are, for example, 150 SCIE/SSCI journals published in Russia, with many "mono-national" Russian papers there (cf. Chankseliani, Lovakov, & Pislyakov, 2021). This should additionally decrease the fractional role of Russian authors in highly cited output.

Why does this problem not occur in journal analysis? Because there are no journal-coauthors and "co-journaled papers". For all other entities, the 1% benchmark is more nuanced than may seem at first glance.

*4.2. Thresholds: general considerations*

But the main problem is, again, with the integer thresholds. To determine the 1%-top in each discipline/year, ESI makes a ranking of papers belonging to this discipline/year from most cited to non-cited. Next, it finds the position of the lower bound of the 1%-top percentile in this ranking. The number of citations received by a paper in this position becomes a threshold for highly cited papers in this discipline/year. Those papers which received not less than the threshold number of citations are highly cited. Those that received less than the threshold, are not highly cited.

The problem is that, according to ESI methodology, the highly cited are not only the 1%-borderline paper and above, but all those that have received *the same number of citations* as the 1%-borderline paper. In other words, often there is no *one borderline paper* as such.

Let us, once again, construct a fictional example to illustrate this. Let us have only 100 papers in the world (or in the given discipline and year). Ninety of them were cited one time each. The remaining 10 were not cited at all. Then, the citation threshold should be equal to 1. Because we cannot prefer one of the cited papers to any other (they received the same number of citations), all 90 papers should be considered highly cited. *Each* of them belongs to the top 1% most cited papers. They are *all* highly cited. This is 90% of the papers' pool!

A somewhat similar example is given by Waltman and Schreiber (2013), who consider 105 publications, 90 of them without citations, 10 with 10 citations each, and five with 20 citations each. They work to resolve the problem of attributing those 10 papers with 10 citations to the top-10% and list different methods of doing so. For example, they mention an "inclusive" approach, when all papers at the threshold are considered highly cited (Bornmann, De Moya Anegón, & Leydesdorff, 2012), an "exclusive" attitude, counting all contentious papers as being not highly cited (Leydesdorff et al., 2011), or a more complex methods of, for example, Van Leeuwen et al. (2003), and Colliander and Ahlgren (2011). They also suggest their own way of handling borderline papers, which contains, so to say, "fuzzy highlycitedness", when papers at the threshold are counted as highly cited in some fraction. This method confers the status of full high citedness on all papers above the borderline. The remaining "space", for highly cited papers to fill the 10% quota, is divided in equal proportions among all borderline papers. As a result, each of them is assigned fractionally to highly cited. The more papers at the threshold, the smaller the proportion of highly cited status each of them receives. This approach always gives an exact total, 10% of top-10% publications, some of them being added up fractionally (Waltman & Schreiber, 2013). It is worth mentioning that this desire, to obtain exactly $n$% of top-$n$% papers, is quite natural. This allows a fair comparison between different subject areas. If we have different real proportions of highly cited papers in different disciplines, then, in some research areas, it is easier to get the distinction mark of publishing a highly cited paper than in the others.

*4.3. Thresholds: suggestions on borderline papers' decision methods with their empirical testing*

We would like to note that all above-mentioned "mechanical", "arithmetical" methods of attributing equally cited papers seem to be a bit artificial. Trying to find *primi inter pares* for borderline papers which received the same number of citations, one should, at first, study some of their substantial characteristics. We suggest three different methods:

(1) To explore the month of publication. Although many databases allow only a year of publication search, a more thorough scrutiny may reveal papers published, for example, in January and December of the same year. Preference should be given to the latter as they had less time to be cited while receiving the same number of citations.[7]

(2) Time distribution of citations received by the papers. If there are more recent citations than those received soon after publication, we may suppose that this borderline paper is on the upswing and probably deserves more to be classified as highly cited.

(3) To explore the citation characteristics of those papers which cite a publication under consideration. This weighted counting may help to fairly select better papers among the set of equally cited publications.[8]

To test these approaches, we collected from InCites all 38,048 mathematics papers (type: Article) published in 2011. We chose mathematics because Waltman and Schreiber (2013) characterized this domain as having especially many papers at the citation threshold. However, it must be said that our problem is easier than theirs. They explored the top-10% papers while we seek the top-1% (according to the highly cited paper definition of ESI). Due to the skewness of citation distributions, the probability of finding papers with the same number of citations at the borderline is lower in our case (suppose if we analyzed papers above median, there would be many more borderline publications at the 50%-level).

---

[7] Of course, this should be used with care and as an inevitable approximation. The real time of paper's appearance may differ from the timestamp on the journal's issue.
[8] One of the more complex methods similar to PaperRank (Du, Bai, & Liu, 2009) or others reviewed by Dunaiski and Visser (2012) may be applied, when citedness of the papers citing those which cite a candidate for highly cited status is taken into account, and so on. Unfortunately, they are rather difficult to implement. Here in a brief demonstration, we simplify the procedure and limit ourselves to highly cited citing papers only (see further).

We sorted papers by number of citations received and found that those placed at the borderline, which is around position #380, have 88 citations. There are nine such papers, with sequence numbers from 377 to 385, that are listed in Table 2. We also manually added two dates to these papers: the date of their first online appearance (being the moment when the final version of the paper became citable) and the month or issue number of their publication.

**Table 2**
Papers in Mathematics (2011, document type: only Articles) at the highly cited threshold. Each paper received 88 citations by now. Sorted in reverse chronological order of "published online" date.

| No | Published online | Published | Article | Journal |
|---|---|---|---|---|
| 1 | 01.11.2011 | Iss. 6 out of 6 | Gross, M., & Siebert, B. From real affine geometry to complex geometry | Annals of Mathematics |
| 2 | 12.08.2011 | November | Bretz, F. et al. Graphical approaches for multiple comparison procedures using weighted Bonferroni, Simes, or parametric tests | Biometrical Journal |
| 3 | 13.07.2011 | Iss. 3 out of 4 | Fujino, O. Fundamental Theorems for the Log Minimal Model Program | Publications of the Research Institute for Mathematical Sciences |
| 4 | 20.05.2011 | August | Ma, W.-X. et al. Wronskian and Grammian solutions to a (3+1)-dimensional generalized KP equation | Applied Mathematics and Computation |
| 5 | 26.03.2011 | May | Ali, M. I. et al. Algebraic structures of soft sets associated with new operations | Computers & Mathematics with Applications |
| 6 | 22.03.2011 | July | Li, X., & Peng, S. Stopping times and related Ito's calculus with G-Brownian motion | Stochastic Processes and Their Applications |
| 7 | 01.02.2011 | May | Engquist, B., & Ying, L. Sweeping Preconditioner for the Helmholtz Equation: Hierarchical Matrix Representation | Communications on Pure and Applied Mathematics |
| 8 | 06.01.2011 | March | Sarkar, B., & Moon, I. An EPQ model with inflation in an imperfect production system | Applied Mathematics and Computation |
| 9 | 28.10.2010 | February | Soner, H. M. et al. Martingale representation theorem for the G-expectation | Stochastic Processes and Their Applications |

We should choose four out of nine papers as highly cited to get 1% from 38,048 = 380 highly cited articles. Papers in Table 2 are sorted in reverse chronological order of the online first date, which helps us to find that papers #1–4 are highly cited, when we use suggested method (1). Note that papers #3 and #6 may be regarded complications if we look at the publication date, because issue 3 of the quarterly journal may appear in July (being the first month of the third quarter). However, when we turn to the online date, we clearly see that paper #3 had an almost 4-month delay compared to #6 and, thus, should be treated as highly cited.[9]

To test suggestion (2), we used a rather rough method and calculated numbers of citations received by these nine papers during 2011–2015 and 2016–2020. Next we took a ratio of the latter value to the former and received coefficients from 2.59 for paper #2 to 0.95 for paper #5. Sorting these results in decreasing order of this ratio gave us a ranking of the papers and allowed us to grant highly cited status to papers #1, 2, 6, and 7.

At last, we tried to approximately assess the citedness of the papers, which, themselves, cite the borderline papers under investigation, suggestion (3). For this purpose, we collected numbers of the highly cited papers among these citing sets. We found that paper #2 was cited by 11 highly cited publications, paper #4 by 7, but the next three papers #7–9 were tied, each of them having received citations from three highly cited documents. Others have two or one such citations.

**Table 3**
Papers from Table 2 and their highly cited status, determined by point-by-point analysis using different indicators: online publication date, regular publication date, ratio of citations received in the last five years to those received in the previous five years, number of highly cited papers among the citing publications. (Citing documents are only from 2011–2020, the ESCI database is excluded.)

| No | Published online | Published | 2020–2016 citations / 2011–2015 citations | Highly cited papers in the set of citing publications |
|---|---|---|---|---|
| 1 | • | • | • | |
| 2 | • | • | • | • |
| 3 | • | • | | |
| 4 | • | • | | • |
| 5 | | | | |
| 6 | | | • | |
| 7 | | | • | (•) |
| 8 | | | | (•) |
| 9 | | | | (•) |

• — determined as highly cited

All results are summarized in Table 3.

After these pilot experiments, we conclude that a detailed chronological approach may be effective in determining which of the borderline papers deserves more to be considered as highly cited. One may use either the online first publication date or "real" date/month of publication or combine both methods. Of course, this thorough procedure should be regarded as "an inevitable evil" in contentious cases. Also, it does not cover such aspects as advance preprint publication which gives an opportunity to cite a work before its final appearance (although, not in its final form). We may add that, in the procedure of defining "hot papers", the ESI database also uses monthly (bi-monthly) division of the publication output for citation analysis (Clarivate, 2021b).

As for other methods, they produce non-unanimous results. Only paper #2 becomes highly cited according to all tested approaches. However, bibliometricians may choose one or another method in line with their particular aims and objectives. Alternatively, one may apply social choice methods to harmonize the results from different techniques as was demonstrated for a set of journal rankings by Subochev et al. (2018).

Of course, other methodologies of finding *primi inter pares* may be suggested. For example, one may take into account the length of each borderline paper, as the more voluminous have higher potential to be cited and have some additional advantage (e.g. Xie et al., 2019; Hasan & Breunig, 2021). Among top-10 most cited papers in physics, for instance, now there are three with more than 1500 pages each. This may be "corrected" in controversial cases. Another possibility is to count citations without self-citations—those when a set of citing paper authors has at least one common person with that of the paper under investigation. The latter is somewhat vulnerable, because excluding all self-citations everywhere would lead to a completely different paper ranking. However, it may be used in a limited way only for equally cited borderline works. What is important is that all such ways of looking at the "borderline problem" try to investigate the *substantial* characteristics of the controversial papers, without recourse to pure *arithmetic* procedures.

### 4.4. Thresholds: ESI database

Now let us turn to empiric data on the highly cited papers. If we take 2011–2018 publications in the ESI database (as data on the most recent papers are subject to outliers and should not be considered), then now, in September 2021, thresholds for the highly cited papers in ESI are in the range from 33 to 493, depending on the discipline/year. For papers in mathematics published in 2018, it is 33. It means that if such a paper is cited 33 times by now, it is "highly cited". If it received 32 citations, it is not highly cited. So, all papers cited 33 or more times are defined as highly cited. As discussed above, it is not necessary that they constitute precisely 1% of the total number of papers. If there is more than one paper with 33 citations, and the number of papers cited 33 or more times does not divide evenly into the total number of mathematics papers, there would be more than 1% of highly cited papers. All papers which received 33 citations are considered as highly cited then. This logic works only in one direction—the number of highly cited papers is 1% *or more* of their total number, not 1% *or less*, because of the ESI "inlcusive" approach to classifying papers as highly cited (see 4.2 above).

**Table 4. ESI: all papers and highly cited papers (the whole database, January 2011–June 2021).**

| Field | ESI total number of papers | Expected number of highly cited papers | ESI number of highly cited papers | Surplus (ESI minus Expected) | **Real % of highly cited papers** |
|---|---|---|---|---|---|
| Agricultural Sciences | 494,887 | 4949 | 4898 | -51 | 0.990 |
| Biology & Biochemistry | 808,554 | 8086 | 8143 | 57 | 1.007 |
| Chemistry | 1,902,618 | 19,026 | 19,263 | 237 | 1.012 |
| Clinical Medicine | 3,133,108 | 31,331 | 31,225 | -106 | 0.997 |
| Computer Science | 448,027 | 4480 | 4503 | 23 | 1.005 |
| Economics & Business | 321,005 | 3210 | 3243 | 33 | 1.010 |
| Engineering | 1,651,456 | 16,515 | 16,464 | -51 | 0.997 |
| Environment/Ecology | 667,445 | 6674 | 6747 | 73 | 1.011 |
| Geosciences | 540,121 | 5401 | 5499 | 98 | 1.018 |
| Immunology | 286,283 | 2863 | 2881 | 18 | 1.006 |
| Materials Science | 1,080,546 | 10,805 | 10,725 | -80 | 0.993 |
| Mathematics | 482,204 | 4822 | 4771 | -51 | 0.989 |
| Microbiology | 236,772 | 2368 | 2392 | 24 | 1.010 |
| Molecular Biology & Genetics | 523,560 | 5236 | 5257 | 21 | 1.004 |
| Multidisciplinary | 25,521 | 255 | 255 | 0 | 0.999 |
| Neuroscience & Behavior | 560,884 | 5609 | 5612 | 3 | 1.001 |
| Pharmacology & Toxicology | 462,351 | 4624 | 4578 | -46 | 0.990 |
| Physics | 1,127,368 | 11,274 | 11,255 | -19 | 0.998 |
| Plant & Animal Science | 813,929 | 8139 | 8011 | -128 | 0.984 |
| Psychiatry/Psychology | 481,833 | 4818 | 4780 | -38 | 0.992 |
| Social Sciences, General | 1,097,043 | 10,970 | 11,018 | 48 | 1.004 |
| Space Science | 158,997 | 1590 | 1588 | -2 | 0.999 |

*Expected number of highly cited papers*: total number of papers / 100, rounded. *ESI number of highly cited papers*: empirical data from the ESI database. *Surplus*: difference between the 4th and the 3rd columns. *Real % of highly cited papers*: ratio of the 4th to the 2nd columns as a percentage.

Let us now look at the empirical result for the whole ESI database at the moment (Table 4). One can see that the number of highly cited papers in the database really differs from (precisely) 1%. But, surprisingly, there are deviations, not only in the positive direction, but also in the negative. The latter is observed for 10 disciplines out of 22 (with one exact match of expected and real numbers). How could that be? The effect described earlier could move the highly cited papers number only above 1% of the total set.

It is possible that the answer is this remark:

> When the calculated threshold is two or fewer for a subject category in a given year, no paper in that category and year receives a Highly Cited designation. This low threshold decision is based on observations that only two citations is small evidence of highly cited paper status. Papers cited at low levels tend to exhibit more volatility with respect to highly cited status over subsequent ESI updates (Clarivate, 2021a).

This could decrease the proportion of the highly cited papers in some disciplines.[10] As a result, two effects act in opposite directions and the number of highly cited papers fluctuates around 1% of the total volume of the set.

## 5. Discussion

Empirical limitations of ranking/baselines methods presented here were demonstrated using one of the most developed and publicly available bibliometric instruments at the moment, the InCites database. In other contexts, the issues discussed may or may not be meaningful. For example, a question of the correct definition of $n$%-top papers is common to InCites' competitor, SciVal database, from Elsevier. This problem is resolved there in a simplified, "inclusive" way too (Elsevier, 2019, pp. 48, 64). However, the SciVal analogue of CNCI, Field-Weighted Citation Impact (FWCI), uses a fractional approach, with harmonic averaging of expected citedness, if a paper has a multidisciplinary attribution (Elsevier, 2019, pp. 47–48, 61). It does not suffer from the predicaments discussed here in Section 3.1 (but does from those of Section 3.2). However, in any context these matters should be scrutinized by a bibliometrician.

The issues found and demonstrated in this paper lead to policy/behavior implications in several aspects:

(a) For the researchers undertaking studies. It is one of the most important messages of the present article, an appeal not to believe in "self-explanatory" terms and always check up the *terminology* by the *methodology* behind it. Ranking is one of the most sound techniques in bibliometrics, still even it has its hidden traps. For example, some of the subject categories in JCR contain no more than nine journals, which, as we have observed in Section 2, leads to 50% difference in numbers of journals across quartiles.

---

[10] Note that this does not completely solve the "90% of highly cited papers problem" from the fictional example above. Ninety cited papers could have received not one, but three citations each, and the threshold would be 3. But this situation is much less probable.

(b) For the educational programs/published works. Avoid the simplistic use of pseudo-self-explanatory definitions, always indicate their limitations. As this paper shows, generally, there are few or no unambiguous terms in bibliometrics. Often the use of such terms is followed with erroneous inferences quasi-intuitively derived from "the names". For example: "These journals are ranked in the first quartile (25 %) of their subject categories. <…> One can expect that 25 % of a researcher's publications have been published in the first quartile" (Bornmann & Marx, 2014).

(c) For database producers. Apart from said above on the use of misleading self-explanatory language in their manuals and webpages, they should, explicitly and loudly, announce if some modifications are made in the methodology they use. For example, the CiteScore procedure change, in 2020, was quite widely demonstrated and explained. This is a positive example. However, five years earlier, the transition from NCI to CNCI in another database, with moving from "ratio of averages" to "average of ratios", was made quite *sotto voce*. We, bibliometricians and scientists, are eager to know all these evolutions in detail and, preferably, in advance.

## 6. Conclusion

This paper is a methodological and instructional piece on the use of well-known terms. We found that *median* is not a median in the strict sense of the term, *percentiles* should be used with reservations and care, *baselines* for the whole world are not unity. These findings are especially crucial when an organization or even a whole country starts to build its research assessment system on rankings/baselines principles (which is, these issues aside, a correct or at least promising approach). The author thinks that further discussion on these issues could contribute to a more sustainable bibliometrics toolbox.

**Acknowledgments**

I would like to thank Ludo Waltman (CWTS) who served as a reviewer for my ISSI-2021 conference submission for his valuable remarks and inspiring ideas, and Pavel Kasyanov (Clarivate) for details of some calculation methodologies in InCites and discussions on the preliminary version of the manuscript. I also thank four reviewers for their valuable comments, and Pete Dick for his review of English.